# A QKD Protocol Extendable to Support Entanglement and Reduce Unauthorized Information Gain by Randomizing the Bases Lists with Key Values and Invalidate Explicit Privacy Amplification

Srinivasan N, Sanjeevakumar C, Kasi Rajan M, Sudarsan L, Venkatesh R
*Department of Information Technology, Madras Institute of Technology, Chennai 600044*
Email : ns@annauniv.edu

## Abstract

*This paper suggests an improvement to the BB84 scheme in Quantum key distribution. The original scheme has its weakness in letting quantifiably more information gain to an eavesdropper during public announcement of unencrypted bases lists. The security of the secret key comes at the expense of the final key length. We aim at exploiting the randomness of preparation (measurement) basis and the bit values encoded (observed), so as to randomize the bases lists before they are communicated over the public channel. A proof of security is given for our scheme and proven that our protocol results in lesser information gain by Eve in comparison with BB84 and its other extensions. Moreover, an analysis is made on the feasibility of our proposal as such and to support entanglement based QKD. The performance of our protocol is compared in terms of the upper and lower bounds on the tolerable bit error rate. We also quantify the information gain (by Eve) mathematically using the familiar approach of the concept of Shannon's entropy. The paper models the attack by Eve in terms of interference in a multi-access quantum channel. Besides, this paper also hints at the invalidation of a separate privacy amplification step in the "prepare-and-measure" protocols in general.*

## l. Introduction

The framework for quantum cryptography essentially lies in the hands of nature [6]. Unless the laws of nature break, the security of the QKD schemes is assured. The most important axioms of quantum mechanics [12, 13, 18] that facilitate this idea are the uncertainty principle, quantum entanglement, teleportation and quantum measurement theory (which is essentially Information gain vs. State disturbance).There are no physical means for gathering information about the identity of a quantum system's state when it is known to be prepared in one of a set of non orthogonal states. This is because of the indistinguishability inherent among such states. This unique feature of quantum phenomena rests on the Hilbert space structure [16] along with the fact that time evolutions for isolated systems are unitary[14, 17, 18]. The uncertainty principle [12, 13, 18] explains the simultaneous measurability of observables (which are mathematicized as operators in Heisenberg mechanics). If two operators commute, they can be measured simultaneously, but non-commuting operators cannot be measured simultaneously. Thus, observables along orthogonal bases cannot be observed at the same time. Conscious observation of one will completely randomize the result in the other. Observation (practically measurement) of a quantum state has its effect in producing a tension between information gain and disturbance. In this paper, we shall quantitatively study the nature of this tension with respect to our protocol (which we refer to as randomization protocol henceforth). Of course, a theoretical description of our protocol is a mathematical idealization. Any real-life quantum cryptographic system is a complex physical system with many degrees of freedom, and is at best an approximation to the ideal protocol. Proving the security of any particular setup is a difficult task, requiring a detailed model of the apparatus. Even a seemingly minor and subtle omission can be fatal to the security of a cryptographic system.

In designing a quantum crypt protocol, it is essential to prove its security under theoretical and practical considerations. Therefore, if one wants to model and analyze the cryptographic security of quantum protocols, one of the most basic questions to be answered is the following. What does it mean for two quantum states to be "close" to each other or "far" apart? That is, we should be concerned with defining and relating various notions of "distance" between two quantum states. Formally a quantum state is nothing more than a square matrix of complex numbers that satisfies a certain set of supplementary properties. Because of this, any of the notions of distance between matrices that can be found in the mathematical literature would do for a quick fix. The only physical means available with which to distinguish two quantum states is that specified by the general notion of a quantum-mechanical measurement. We include, in this paper an analysis of our protocol by considering different distinguishability measures.

All existing QKD schemes are variations to the fundamental BB84 protocol [2, 8]. In the BB84 protocol, Alice sends a qubit (i.e., a quantum bit or a two-level quantum system) in one of four states to Bob. The states $|0\rangle$ and $|+\rangle = (|0\rangle$ and $|1\rangle)/\sqrt{2}$ and represent the classical bit 0, while the states and $|1\rangle$ and $|-\rangle = (|0\rangle - |+\rangle)/\sqrt{2}$ represent the bit 1. Alice chooses one of these four states uniformly at random, and sends it to Bob, who chooses randomly to measure in either the |0>, |1> basis (the "Z" basis) or the $|+\rangle, |-\rangle$ basis (the "X" basis). Then, Alice and Bob announce the basis each of them used for each state (but not the actual state sent or measured in that basis), and discard any bits for which they used different bases. The remaining bits form the raw key, which will be processed some more to produce the final key.

In the scheme as proposed by Hannes R Bohm et al [2], in addition to the lists of BB84, Alice and Bob both possess a preshared secret (randomizer) that is split into two parts, S-Alice and S-Bob. The information which basis was used during each individual measurement is encrypted before it is sent over the classical channel using the shared key. This is done by applying logic XOR between the list of bases and a part of the shared secret. This encryption of the encoding and measurement bases renders it impossible for a third party to correctly sift measurement results obtained from eavesdropping on the quantum channel. For the protocol to be secure it is mandatory that Alice and Bob use different parts of the shared secret and that for successive runs of the protocol, the secrecy of the shared secret has to be continuously refreshed. Our proposal has its roots in this scheme with the added benefit of invalidating explicit privacy amplification step and the randomizer need not be refreshed for every run of the protocol.

The QKD scheme without public announcement of basis has been proposed by Won Young Hwang et al [1]. In this protocol, Alice and Bob share by any method (BB84 scheme) some secure binary random sequence that is known to nobody. This random sequence is to be used to determine the encoding basis u and u'. Alice (Bob) encodes (performs spin-measurement) on the basis z and x when it is 0 and 1, respectively. For example, when the bases random sequence is 0, 1, 1, 0, 1.....and the signal random sequence that Alice wants to send to Bob is 1, 0, 1, 0, 1..... Then she sends to him the following quantum carriers $|z-\rangle$, $|x+\rangle$, $|x-\rangle$, $|z+\rangle$, $|x-\rangle$,...... Since Alice and Bob have common random sequence, there will be perfect correlation between them unless the quantum carriers were perturbed by Eve or noise. Eve is naturally prevented from knowing about the encoding bases, since she does not know the bases sequence. As we see, public announcement of bases is not needed in the proposed scheme. However, the scheme can only be useful if it is possible to use safely the bases random sequence repeatedly. If this is not the case, Alice and Bob have to consume the same length of random sequences to obtain some length of new random sequences. Fortunately, quantum mechanical laws enable the bases random sequences to be used repeatedly enough.

In the Inamori's Protocol [5] Alice and Bob are assumed to share initially a random string and the goal of QKD is to extend this string. Alice and Bob also choose a classical error-correcting code C1. Alice sends Bob a sequence of single photons as in either BB84 or the six-state scheme. They throw away all polarization data that are prepared in different bases and keep only the ones that are prepared in the same bases. They randomly select m of those pairs and perform a refined data analysis to find out the error rate of the various bases. Alice measures the remaining N-m =s particles to generate a random string. Being a random string, it generally has nontrivial error syndrome when regarded as a corrupted state of the codeword of C1. Alice transmits that error syndrome in an encrypted form to Bob. This is done by using a one-time pad encryption with (part of) the common string they initially share as the key. Bob corrects his error to recover the string. They discard all the bits where they disagree and keep only the ones where they agree. Finally, Alice and Bob perform privacy amplification on the remaining string to generate a secure string.

One important class of protocols is the Entanglement Purification Protocol schemes for QKD [5, 8]. This is a generalized scheme and can be reduced to the standard BB84 and its extensions.

Suppose Alice and Bob are connected by a noisy quantum channel (and perhaps also a noiseless classical channel). Entanglement purification provides a way of using the noisy quantum channel to simulate a noiseless one. More concretely, suppose Alice creates N EPR pairs and sends half of each pair to Bob. If the quantum communication channel between Alice and Bob is noisy (but stationary and memory-less), then Alice and Bob will share imperfect EPR pairs, each in the state P. They may attempt to apply local operations (including preparation of ancillary qubits, local unitary transformations, and measurements) and classical communications (LOCCs) to purify the N imperfect EPR pairs into a smaller number n, EPR pairs of high fidelity. This process is called an EPP. One way to classify EPPs is in terms of what type of classical communications they require: 'EPPs that can be implemented with only one-way classical communications from Alice to Bob, known as 1-EPPs' and 'EPPs requiring two-way classical communications, known as 2-EPPs'.

Typically, a 1-EPP will consist of Alice measuring a series of commuting operators and sending the measurement result to Bob. Bob will then measure the same operators on his qubits. If there is no noise in the channel, Bob will get the same results as Alice, but of course when noise is present, some of the results will differ. From the algebraic structure of the list of operators measured, Bob can deduce the location and nature of the errors and correct them. Unfortunately, the process of measuring EPR pairs will have destroyed some of them, so the resulting state consists of fewer EPR pairs than Alice sent.

2-EPPs can be potentially more complex, but frequently have a similar structure. Again, Alice and Bob measure a set of identical operators. Then they compare their results, discard some EPR pairs, and together select a new set of operators to measure. An essential feature of a 2-EPP is that the subsequent choice of measurement operators may depend on the outcomes of previous measurements. This process continues for a while until the remaining EPR pairs have a low enough error rate for a 1-EPP to succeed. Then, a 1-EPP is applied.

The proof of security of this class has been worked out already and its result is used for proving that our protocol is secure.

In section II of our paper, the preliminary mathematical results and the basic notations used in the analysis of security considerations and information gain with respect to our protocol are discussed. In section III, we present a theoretical description of our protocol. In section IV, we analyze its security extensively and model all types of attacks by an eavesdropper. We quantify the information gain in the subsequent section. The final section includes a theoretical consideration of incorporating randomization into entanglement based QKD.

## 2. Mathematical Notations and Definitions

This part enables the reader to understand the mathematical concepts and their notations used in quantum information processing [15, 19]. Much of the math given here will enable the reader to visualize the mathematical underpinnings given in our paper in support of our protocol.

### 2.1. Operators in Quantum Theory with respect to Quantum Cryptography

Density operators are used for various purposes in quantum theory, and their significance depends somewhat on how they are used. Usually they function as pre-probabilities, i.e., they are devices for calculating quantum probabilities. Think of a density operator as something like a probability distribution in classical physics. The most common uses of a density operator are: to provide partial information about a system, in analogy with a probability distribution in classical physics (the operator $e^{-H/kT}/z$ in statistical mechanics is an example), and to describe a subsystem of a total system made up of two (or more) parts.

A *density $\rho$ matrix* is a $N \times N$ matrix with unit trace that is Hermitian $(i.e., \rho = \rho^{\dagger})$ and positive semi-definite $i.e., \langle \Psi | \rho | \psi \rangle \geq 0$ for all $\Psi \in H$

### 2.1.1. Quantum Ensembles.
Suppose Alice is instructed to generate the quantum state $|\Psi_j\rangle$ with probability $P_j$, and ship it to Bob in a closed box which is carefully isolated from the environment and keeps the quantum system isolated, so the dynamics is trivial.

One speaks of a quantum ensemble $\{P_j, |\Psi_j\rangle\}$. Here the $|\Psi_j\rangle$ are assumed to be normalized, but they don't have to be orthogonal to each other. E.g., Alice sometimes prepares a qubit in state $|o\rangle$, sometimes in state $|+\rangle$, etc. The idea of an ensemble can be used even if one is only talking about one event, since it is just a method of visualizing probabilities. The unopened box is in front of Bob, who wonders whether this system "is in the state $|s\rangle$," by which he means either in $|s\rangle$ or some state orthogonal to $|s\rangle$; the corresponding sample space is $\{[S], I - [S]\}$, where $[S] = |s\rangle\langle s|$. Bob reasons as follows. If Alice prepared $|\Psi_j\rangle$, then the desired probability is given by the Born rule, and is $P(s|j) = \langle \Psi_j |[s]| \Psi_j \rangle$.

But he doesn't know which state Alice prepared; all he knows is that the probability of her preparing $|\Psi_j\rangle$ is $P_j$. Therefore he assigns to the system in front of him a probability

$$P(s) = \sum_j P(s|j) p_j = \sum_j |\langle s|\psi_j\rangle|^2 p_j$$

$$P(s) = \sum_j P(s|j) p_j = \sum_j |\langle s|\psi_j\rangle|^2 p_j$$

where

$$\rho := \sum_j p_j |\Psi_j\rangle\langle\Psi_j| = \sum_j p_j [\Psi_j]$$

is the density operator associated with the ensemble $\{P_j, |\Psi_j\rangle\}$, and $T_r$ means the trace. The role of density operators, as in this example, is to serve as convenient tools for calculating probabilities. A density operator provides partial information about a system in circumstances where more information is potentially available, the same as with a probability.

In particular, in the case of a state prepared by Alice, more information than provided by ρ is at least potentially available. Bob could ask Alice, who knows which state $|\Psi_j\rangle$ she prepared. Or he could go ahead and measure the system in front of him to see if it is or is not in the state $|s\rangle$. Unless it is a pure state the same density operator can be derived from many different ensembles $\{P_j, |\Psi_j\rangle\}$. In addition to ensembles of kets, or pure states, it is sometimes convenient to use ensembles of mixed states, $\{P_j, \rho_j\}$, with a density operator for the ensemble given by $\rho = \sum_j P_j \rho_j$ with, again, nonnegative probabilities $\{P_j\}$ that sum to 1.

Two different ensembles with the same density matrix are indistinguishable as far as an observer is concerned; when this is the case, there exists no measurement that can allow the observer a decision between the ensembles with probability of success better than chance.

Now consider the following preparation of a quantum system: A flips a fair coin and, depending upon the outcome, sends one of two different pure states $|\Psi_0\rangle$ or $|\Psi_1\rangle$ to B. Then the "pureness" of the quantum state is "diluted" by the classical uncertainty about the resulting coin flip. In this case, no deterministic fine-grained measurement generally exists for identifying A's exact preparation, and the quantum state is said to be a *mixed* state. B's knowledge of the system—that is, the source from which he draws his predictions about any potential measurement outcomes—can now no longer be represented by a vector in a Hilbert space. Rather, it must be described by the density operator from a statistical average of the projectors associated with A's possible fine-grained preparations.

Now, we describe how to compute the probability of a certain measurement result from the density matrix. Mathematically speaking, a density matrix $\rho$ can be regarded as an object to which we can apply another operator $E_x$ to obtain a probability. In particular, taking the trace of the product of the two matrices yields the probability that the measurement result is given that the state was $\rho$, i.e.,

$$\Prob[result = x | state = \rho] = Tr(\rho E_x)$$

Here x the serves as a label, connecting the operator $E_x$ and the outcome x, but otherwise has no specific physical meaning. (This formula may help the reader understand the designation "density operator": it is required in order to obtain a probability density function for the possible measurement outcomes.)

## 2.2. Some Important Properties of the Operators

1. A density operator (density matrix) is a positive operator on the quantum Hilbert space with unit trace.
2. An operator is positive if it is (i) Hermitian, and (ii) none of its (necessarily real) eigen values is less than zero (positive semi-definite or nonnegative semi-definite). More concretely, R is positive if and only if $\langle\Psi|R|\Psi\rangle$ is real and $\langle\Psi|R|\Psi\rangle \geq 0$ for every $|\Psi\rangle$ in the Hilbert space. The trace of an operator Q, written Tr(Q) or tr(Q), is $Tr(Q) = \sum_j \langle j|Q|j\rangle$ using any orthonormal basis $\{|j\rangle\}$. The result is independent of which orthonormal basis one uses.
3. The eigen values $\{\lambda_j\}$ of a density operator ρ are nonnegative, and since they sum to 1 they must all lie between 0 and 1 (like probabilities). Two cases are distinguished: if one of the eigenvalues is 1 and the rest are 0, ρ is (or refers to or corresponds to) a pure state. In all other cases it is a mixed state. In the case of a pure state, one can always write ρ in the form $|\Psi\rangle\langle\Psi|$ for some normalized ket $|\Psi\rangle$.

Like any Hermitian operator, ρ can be expanded in terms of projectors on its eigenvectors,

$$\rho = \sum_j \lambda_j |a_j\rangle\langle a_j|$$

where the $\{|a_j\rangle\}$ form an orthonormal basis. The rank of ρ is the number of nonzero eigenvalues (degenerate eigenvalues counted more than once).

### 2.3. Density Operator for a Subsystem

Consider a system consisting of two subsystems A and B, with a Hilbert space which is a tensor product $H_a \otimes H_b$. Suppose the state of the total system at some time, thought of as a pre-probability (e.g., obtained by integrating Schrödinger's equation from a state at an earlier time) is $|\Psi\rangle$. Then we can use it to calculate probabilities in any decomposition of the identity that interests us. In particular, we may be interested in properties of the system A corresponding to a decomposition of its identity

$$I_a = \sum_j [a_j] = \sum_j |a_j\rangle\langle a_j|$$

We can write the probability as:

$$\Pr(a_j) = \langle\Psi|[a_j]|\Psi\rangle = Tr[([a_j]\otimes I)|\Psi\rangle\langle\Psi|] = Tr_a(a_j\rho)$$

where

$$\rho = Tr_b(|\Psi\rangle\langle\Psi|)$$

is called the reduced density operator for subsystem A.

In the course of this paper, we describe other relevant mathematical results that characterize the development of the randomization protocol.

## 3. Theoretical Description of the Randomization Protocol

For our present discussion let us idealize a noiseless quantum channel. The randomization protocol that we propose exploits the randomness in the preparation (measurement) basis and bit values sent (observed). We base our protocol in the fact that randomizing the basis-list (which contains classical bit values representing the choice of basis) before being publicly announced will consequently reduce information gain by an eavesdropper on the final sifted key. We understand that the scheme, as suggested by Hannes R. Bohm [2], randomizes the basis-list. However, the shared secret S used to XOR with the original basis-list remains the same. In subsequent runs of the protocol, Eve can obtain fairly better information as to bit values in the shared secret. Note that knowledge on the shared secret S that has been used during basis reconciliation is equivalent to knowledge on sifting function f that was used to create the sifted key.

We can extend the same scheme to hide the shared secret S through quantum data hiding [9]. This implies that S can be dynamically generated by one of the parties and communicated to the other for each run of the protocol. The quantum data hiding scheme is secure if Alice and Bob cannot communicate quantum states and do not share prior quantum entanglement.

This is exactly the situation we are in. We need to communicate beforehand the secrets bit string S. It has been found that even states without quantum entanglement can exhibit properties of nonlocality.

We now present the phases involved in our randomization protocol.

1. (Quantum transmission). The BB84 quantum transmission is executed. Alice encodes random bit values in random basis and sends the qubits to Bob via the quantum channel. Bob measures the qubits under a random choice of measurement basis. Since, we have assumed a noiseless channel, the quantum states prepared by Alice remain undisturbed. Thus Bob will observe the same bit values as encoded for the same choice of basis. In the most trivial case, the bit string encoded and observed is assumed to be the same. This is true for bit values that have been prepared and measured in same basis (since the channel is noiseless). For different choice of basis the probability that the measured value being same as the encoded value is 1/2.

2. (XOR operation). For the trivial case, the basis list of Alice A is XORed with the random bit string K (which she encodes) to produce a new list $N_a$. The same operation is performed by Bob at the other end (his bases list B XORed with the observed bit string K to produce $N_b$).

3. (Announcement of randomized bases lists). Alice and Bob publicly announce their respective randomized lists $N_a$ and $N_b$

4. (Retrieval of bases). Alice performs an XOR between $N_b$ and K to obtain the bases list of Bob. At the other side of the link, Bob performs XOR between Na and K to obtain Alice's basis list. Thus, the choice of bases of each party has been announced secretly to the other.

5. A common sifted key is generated by discarding the bit values for which the choice of bases was not same.

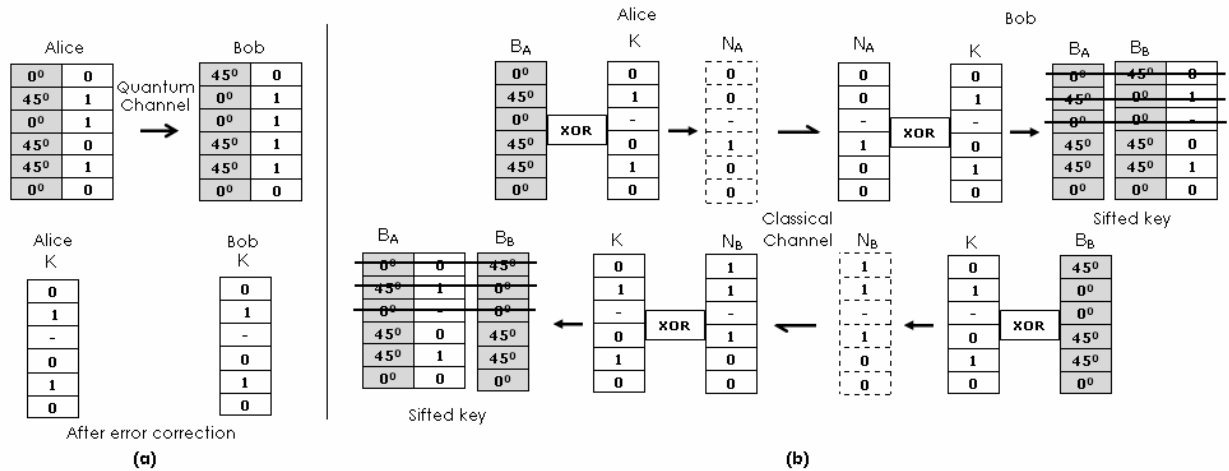

Figure 1. (a) Quantum Channel Transmission and after error correction (b) A comprehensive pictorial flow of our Randomization protocol

There is considerable change in our protocol for the more practical considerations. This involves taking into account multi-access interference, signal attenuation and random noise (eavesdropping). Under this condition, after the initial step of quantum transmission, error estimation and error correction are performed over the raw key (rather on the sifted key as in the original BB84).

Let's suppose that the raw key is of length p. Alice announces a subset of positions of size k and the bit values for those positions in the raw key. Bob also returns the bits he received in those positions. Both compute the observed error-rate e and accept the quantum transmission if e <= $e_{max}$ as set initially by Alice. They remove the k bits announced from the raw key. If e > $e_{max}$ then they abort the current run of the protocol. Error correction subsequently follows. Alice selects L random subsets $X_1...X_L$ of positions and announces $X_i$ (i=1 to L) together with the parity of all bits in $X_i$. Bob compares the parity bits announced by Alice to the one he gets with his bits and tells Alice whether they are all the same. If some parity does not match, then Alice and Bob abort. Finally, we have an error free raw key as in Figure 1. (a). Both parties now have the same key string K. Now the protocol may terminate here. But, in order to incorporate implicit privacy amplification steps 3, 4, 5 are carried out as in Figure 1. (b). This protocol, as we see, invalidates the necessity for explicit privacy amplification procedure.

### 3.1. Feasibility of the Randomization Protocol

The feasibility of the randomization protocol lies in effective quantum error correction. Our protocol requires QEC to be performed over the entire raw key. The error correction scheme has to be exact in order to produce the key and randomize the bases lists on either side of the channel.

It is an immediate result of the no-cloning theorem that no quantum error-correcting code of length n can fix n/2 erasures because that would imply that we could reconstruct two copies of an encoded quantum state from two halves of the full codeword. This statement is valid regardless of the dimension of the coding Hilbert space. Another well known result from the theory of quantum error correction is that a length n code can fix t arbitrary single position errors if and only if it can fix 2t erasure errors. This follows immediately from the quantum error-correction conditions

$$\langle \Psi_i | E_a^\dagger E_b | \Psi_j \rangle = C_{ab} \delta_{ij}$$

(for basis encoded states $\{|\Psi_i\rangle\}$ and correctable errors $\{E_a\}$) and implies that no QECC of length n can fix more than n=4 arbitrary errors, regardless of the dimension of the coding and encoded Hilbert spaces.

Now, this turns out to be a genuine restriction of our scheme. QEC in the BB84 scheme is applied only on bit values measured in the same bases. This implies that the QEC applies on key whose length is lesser in comparison with our scheme and thereby, more robust. In the worst case, the bit values encoded and observed differ in all bit positions. This renders our scheme unusable with Alice and Bob immediately aborting the protocol operation. The best case where there are no bit flip errors irrespective of the bases, matches the trivial case we had discussed earlier. In the average case, we might expect exactly n/2 errors for a raw key of length n. For Alice and Bob to continue, QEC has to correct n/2 bits exactly. But, there is no such exact error-correction scheme that will do this unless we retort to an approximate QEC whose fidelity is exponentially closer to 1 [21].

## 4. Modeling of Attacks and Proof of Security

In this section, we present detailed security considerations of our protocol. We model all types of attacks or eavesdropping strategies by Eve and subsequently prove that our protocol remains secure. Our protocol is essentially BB84 with the public announcement of the bases encrypted.

Our protocol improvises upon the one suggested by Hannes R.Bohm [2] in that the shared secret S in the original scheme is taken to be the shared error-corrected key string. For each run of our protocol, the key string is new and its secrecy completely refreshed owing to the randomness of key string encoded (observed).

### 4.1. Security Requirements of the Protocol

Naively, one might consider a security requirement of the form $I_{eve} < \delta(n)$, where $I_{eve}$ represents eavesdropper's mutual information with the final key and n is is the length of the final key. However, such a definition of security is too weak, since it allows Eve to learn a few bits of a long message. For instance, the eavesdropper may know something about the structure of the message that Alice is going to send to Bob. Another naive definition of security would be to require that $I_{eve} < e^{-kn}$ for any eavesdropping strategy. Unfortunately, such a definition of security is too strong to be achievable. For instance, Eve can simply replace the signal prepared by Alice by sending Bob some signals with specific polarizations prepared by her. Such an eavesdropping attack is highly unlikely to pass the verification test (by producing a small error rate). However, in the unlikely event that it does pass the verification test, Eve will have perfect information on the key shared between Alice and Bob, thus violating the security requirement $I_{eve} < e^{-kn}$. For this paper, however, we simply use the following definition:

A QKD protocol to generate key bits is *correct* if, for any strategy used by Eve, either Alice or Bob will abort with high probability or, with high probability, Alice and Bob will agree on a final key which is chosen nearly uniformly at random. The protocol is *secure* if, for any strategy used by Eve, either Alice or Bob will abort with high probability or Eve's information about the key will be at most $e^{-s}$ for some security parameter *s*. In all cases, "with high probability" means with probability at least $1 - e^{-r}$ for some security parameter *r*. The resources required for the implementation of a QKD scheme must be at most polynomial in r and s. For simplicity, in what follows, we will consider the case where r = s = n (key length) and call it simply the security parameter.

### 4.2. Types of Attacks and Eavesdropping Strategies

All of the protocol operations we consider will take place over a noisy quantum channel, even when there is no eavesdropper present. We shall be primarily interested in a special class of quantum channels known as Pauli channels [5, 8]. From the perspective of Alice and Bob, noise in the channel could have been caused by an eavesdropper Eve. We will need to consider two types of eavesdropping strategy by Eve. The first strategy, the joint attack, is the most general attack allowed by quantum mechanics. In a *joint attack* by Eve, Eve has a quantum computer. She takes all quantum signals sent by Alice and performs an arbitrary unitary transformation involving those signals, adding any additional ancilla qubits she cares to use. She keeps any part of the system she desires and transmits the remainder to Bob. She listens to the public discussion (for error correction/detection and privacy amplification) between Alice and Bob before finally deciding on the measurement operator on her part of the system. The joint attack allows Eve to perform any quantum operation on the qubits transmitted by Alice. For the security proof, we shall also consider a Pauli attack. *A Pauli attack* by Eve is a joint attack where the final operation performed on the transmitted qubits is a general Pauli channel.

Our protocol remains secure under a probabilistic clone/resend attack strategy [11]. This attack has the following description: Eve employs such a cloning machine. She may clone every polarization quantum states sent by Alice and then resend Bob a new one corresponding to her own result. The input state sent by Alice is one of the four quantum states *so* that the sets of input states can be easily enumerated. The probability of each set occurs can be figured out in terms of the corresponding probability of quantum state sent by Alice. The input states sets can be categorized into three divisions according to the type of 'overlapping' of two input states. If two input states are orthogonal, the maximum cloning efficiency approaches to 1,

$$\eta \leq \frac{1}{1+\langle \Psi_0 | \Psi_1 \rangle} = \frac{1}{1+\langle \leftrightarrow | \updownarrow \rangle} = \frac{1}{1+\langle + | - \rangle} = 1$$

When two input states are non-orthogonal, the

maximum cloning efficiencies can be calculated as following:

$$\eta \leq \frac{1}{1+\langle -|\updownarrow \rangle} = \frac{1}{1+\langle +|\updownarrow \rangle} = 2-\sqrt{2} = 0.5858$$

$$\eta \leq \frac{1}{1+\langle -|\leftrightarrow \rangle} = \frac{1}{1+\langle +|\leftrightarrow \rangle} = 2-\sqrt{2} = 0.5858$$

Incase of two input states are the same, the maximum cloning efficiency becomes 0.5

$$\eta \leq \frac{1}{1+\langle \updownarrow|\updownarrow \rangle} = \frac{1}{1+\langle \leftrightarrow|\leftrightarrow \rangle} = \frac{1}{1+\langle +|+ \rangle} = \frac{1}{1+\langle -|- \rangle} = 0.5$$

Given the probability each input states set occurs and corresponding maximum cloning efficiencies, the average of cloning efficiency can be calculated as following:

$$\bar{\eta} = 2 \times \frac{1}{8} \times 1 + 4 \times \frac{1}{8} \times 0.5858 + 4 \times \frac{1}{16} \times 0.5 = 0.6679$$

It means that Eve can clone 66.79% quantum states sent by Alice when she use the probabilistic cloning machine to eavesdrop. So Eve is able to resend Bob a correct qubit in 66.79% cases, while Alice and Bob would not notice her intervention. However, in the rest 33.21% cases, Eve can't clone states correctly; Alice and Bob may discover her intervention in about half of these cases (l6.6%), due to getting uncorrelated results. Consequently, the information that Eve could get in probabilistic clone/resend attacks fashion is 83.4% and the QBER of the sifted key for Alice and Bob is 16.6%.

### 4.3. Proof of Security

Our scheme belongs to the class of 'Prepare-and-Measure' protocols [5]. We here prove that this class of protocols are secure. We have already presented an account of the EPP (1-way and 2-way) schemes [5]. In the EPP scheme, Alice creates N EPR pairs and sends half of each to Bob. Alice and Bob then test the error rates in the X and Z bases on a randomly chosen subset of m pairs. If the error rate is too high, they abort; otherwise, they perform an EPP on the remaining N-m pairs. Finally, they measure (in the Z basis) each of the EPR pairs left after C, producing a shared random key about which, they hope, Eve has essentially no information. The security of our protocol can be derived from the proof of security of EPP schemes. The reduction to a "prepare-and-measure" protocol is done as a series of modifications to the EPP protocol to produce equivalent protocols. The main insight is that the X-type measurements do not actually affect the final QKD protocol, and therefore are not needed. The X-type measurements give the error syndrome for phase (Z) errors, which do not affect the value of the final key. Instead, Z errors represent information Eve has gained about the key. The phase information thus must be delocalized, but need not actually be corrected. The upshot is that Alice and Bob need not actually measure the X-type operators and can therefore manage without a quantum computer. Our initial goal is to manipulate the EPP protocol to make this clear. The X-type measurements do not, however, disappear completely: instead, they become privacy amplification. For the first step, we modify the EPP to put it in a standard form. Because it is a CSS-like 1-EPP, and each operator being measured is either X-type or -Z type. The operators all commute, and do not depend on the outcome of earlier measurements, so we can reorder them to put all of the Z-type measurements before all of the X-type measurements. Now we have an EPP consisting of a series of Z-type measurements, followed by a series of X-type measurements, followed by CNOTs and Pauli operations (which we can represent as I, X, and/or Z on each qubit). Then Alice and Bob measure all qubits in the Z basis. As a second step, we can move all X Pauli operations to before the X-type measurements, since they commute with each other. Moreover, if Alice and Bob are simply going to measure a qubit in the Z basis, there is no point in first performing a Z phase-shift operation, since it will not affect at all the distribution of outcomes of the measurement. We now have an EPP protocol consisting of Z-type measurements, followed by X Pauli gates, followed by X-type measurements, followed by a sequence of CNOT gates which does not depend on the measurement outcomes.

But nothing in the current version of the protocol depends on the outcomes of the X-type measurements, so those measurements are useless. We might as well drop them. Furthermore, X Pauli operations and CNOT gates are just classical operations, so we might as well wait to do them until after the Z basis measurement, which converts the qubits into classical bits. What's more, it is redundant to perform Z-type measurements followed by measurement of for each qubit. We can deduce with complete accuracy the outcome of each Z-type measurement from the outcomes of the measurements on individual qubits. For instance, if a sequence of three bits is measured to have the value 101, then we know that measurement of Z*Z*Z will give the result +1, as the parity of the three bits is even. Thus, we are left with the following protocol: Alice prepares a number of EPR pairs, and sends half of each to Bob. She and Bob each perform the correction rotation (I or H for the two-basis scheme; I, T, TT or for the three-basis scheme), then measure each qubit in the Z basis. They use some of the results to test the error rate, and on the rest they perform some classical gates derived from the original EPP.

When the EPP is based on a CSS code, the Z-type operators correspond to the parity checks of a classical error-correcting code C1, and the X-type operators correspond to the parity checks of another classical code $C2^\perp$, with $C2^\perp$ subset of C1. The quantum codewords of the CSS code are super positions of all classical codewords from the cosets of $C2^\perp$ in C1. Measuring the Z-type operators, therefore, corresponds to determining the error syndrome for C1, whereas measuring the X-type operators determines the error syndrome for C2. The usual 1-EPP protocol for correcting errors is for Bob to compute the difference, in both bases, between Alice's syndrome and his syndrome, and then to perform a Pauli operation to give his states the same syndromes as Alice's state. That is, Alice and Bob now each have a superposition over the same coset of C2' within the same coset of C1 (or rather, they have an entangled state, a superposition over all possible shared cosets for a given pair of syndromes). The decoding procedure then determines which coset $C2^\perp$ of they share and use that as the final decoded state.

More concretely, we can describe the classical procedure as follows: For the error correction stage, Alice computes and announces the parity checks for the code C1. Bob subtracts his error syndrome from Alice's and flips bits (according to the optimal error-correction procedure) to produce a state with relative error syndrome; that is, he should now have the same string as Alice. Then Alice and Bob perform privacy amplification: they compute the parity checks of C2' (i.e., they multiply by the *generator* matrix of C2) and use those as their final secret key bits. There is one final step to convert the protocol to a "prepare-and-measure" protocol. Instead of preparing N qubits and sending them to Bob, Alice prepares 2N+e (for BB84) or 3N+e (for the six-state scheme). Instead of waiting for Alice to announce which rotation she has performed (I, H, T or TT), Bob simply chooses one at random. Instead of rotating and then measuring in the Z basis, Bob simply measures in the X, Y, or Z basis, depending on which rotation he chose. Then Alice and Bob announce their bases, and discard those bits for which they measured different bases. With high probability, there will be at least N remaining bits. Alice and Bob perform the error test on m of them, and do error correction and privacy amplification on the remaining N-m. Since the discarded bits are just meaningless noise, they do not affect the security of the resulting "prepare-and-measure" protocol. The only difference is that security must now be measured in terms of the remaining bits rather than the original number of qubits sent. When we begin with a two-basis scheme, we end up with BB84; when we begin with a three-basis scheme, we end up with the six-state protocol.

From the above discussion, we come to the point where the proof of security of our randomization protocol is implied.

## 5. Quantifying Information Gain

Suppose Eve obtains a quantum system secretly prepared in one of two nonorthogonal pure quantum states. Quantum theory dictates that there is no measurement she can use to certify which of the two states was actually prepared. This is well known. A simple, but less recognized, corollary is that no interaction used for performing such an information-gathering measurement can leave both states unchanged in the process. If Eve could completely regenerate the unknown quantum state after measurement, then-by making further nondisturbing information-gathering measurements on it-she would eventually be able to infer the state's identity after all. In this section, we present the quantification of information gain and disturbance due to Eve's intervention in the quantum channel. All of the above attacks have been taken into consideration. We shall also see the quantized nature of Eve's information with respect to the attacks on the classical channel. We extensively make of use of the results derived by Christopher A. Fuchs [4, 10].

### 5.1. The Scenario

Alice randomly prepares a quantum system to be in either a state $\hat{\rho}_0$ or a state $\hat{\rho}_1$. These states, in the most general setting, are described by density operators on an N-dimensional Hilbert space for some N; there is no restriction that they be pure states, orthogonal, or commuting for that matter. After the preparation, the quantum system is passed into a "black box" where it may be probed Eve in any way allowed by the laws of quantum mechanics. That is to say, Eve may allow the system to interact with an auxiliary system leaving the probe ultimately in one of two states $\hat{\rho}_{E_0}$ or $\hat{\rho}_{E_1}$ so that after the systems have decoupled, she may perform quantum mechanical measurements on the probe itself. Because the outcome statistics of the measurement will then be conditioned upon the quantum state that went into the black box, the measurement may provide Eve with some information about the quantum state and may even provide her a basis on which to make an inference as to the state's identity. Eve now has the potential to gather information about identity of

Alice's preparation, via the alternate states of her probe. Meanwhile the states $\hat{\rho}_0$ and $\hat{\rho}_1$ no longer form valid descriptions of Alice's system because it will have become entangled with Eve's ancilla. We now have the following list of considerations:
• a concise account of all probes and interactions that Eve may use to obtain evidence about the identity of the state.
• a convenient description of the most general kind of quantum measurement she may then perform on her probe.
• a measure of the information or inference power provided by any given measurement.
• a good notion by which to measure the distinguishability of mixed quantum states and a measure of disturbance based on it.

## 5.2. Mathematical Description of the Quantitative Analysis

The most general notion of measurement allowed within quantum mechanics is the POVM (short for Positive Operator-Valued Measure). This type results in classical information.

Lets assume the following notations:
$\hat{\rho}_s, s = 0,1$ state of the system prepared by Alice and is the density operator on Hilbert space $H_A$

$\hat{U}$ unitary interactions

$\hat{\sigma}$ some standard density operator on probe's (Eve) Hilbert space $H_E$

$tr_E$ represents a partial trace over Eve's probe.

$tr_A$ represents a partial trace over Alice's system.

$P_s(b)$ is the probability of the various outcomes of the measurement by Eve

$\{\hat{E}_b\}$ POVM (set of operators satisfying $\langle \Psi | \hat{E}_b | \Psi \rangle > 0$ and $\sum_b \hat{E}_b = \hat{I}_e$

$\otimes$ Tensor product

The POVM formulation of a measurement is particularly convenient for optimization problems because not only can POVMs be derived from specific measurement models but, conversely, any set of operators $\{\hat{E}_b\}$ satisfying the definition of a POVM can be identified with a measurement procedure [7].

This gives an easy algebraic characterization of all possible measurements. We have the following result
$$P_s(b) = tr(\hat{\rho}_s^E \hat{E}_b)$$
Let $\varepsilon = \langle E_1, ...., E_m \rangle$ be a collection (ordered set) of operators such that 1) $E_x$ are all positive semi-definite operators, and 2) $\sum_x E_x = ID$, where $ID$ is the identity operator. Such a collection specifies a *Positive Operator-Valued Measure* (POVM) and corresponds to the most general type of measurement that can be performed on a quantum system.

Applying a POVM to a system whose state is described by a density matrix results in a probability distribution according to
$$\Pr ob[result = x \mid state = \rho] = Tr(\rho E_x)$$

There are several ways by which to quantify the "information" that Eve gathers about the identity of Alice's preparation. We shall consider just the most suitable information and distinguishability measure. This method, in one sense or another, describe how much "information" can be gained about the identity of a quantum state. Also this method is truly information theoretic in the sense of Shannon's information theory.

Suppose Eve measures a POVM { $\hat{E}_b$ } on $H_E$. If Alice prepared $\hat{\rho}_0$, the outcomes of Eve's measurement will occur with probabilities $P_0(b) = tr(\hat{\rho}_0^E \hat{E}_b)$; if Alice prepared $\hat{\rho}_1$, the outcomes will occur with probabilities $P_1(b) = tr(\hat{\rho}_1^E \hat{E}_b)$.

Given this measurement on Eve's part, the extent to which Alice's preparations can be distinguished is exactly the extent to which the probability distributions $P_0(b)$ and $P_1(b)$ can be distinguished.

A very simple measure of the distinguishability of $P_0(b)$ and $P_1(b)$ is the statistical overlap between these distributions
$$B(p_0, p_1) = \sum_b \sqrt{p_0(b)} \sqrt{p_1(b)}$$

When there is no overlap between the distributions they can certainly be distinguished completely. Alternatively, the overlap is unity if and only if the distributions are identical and cannot be distinguished at all. Such a measure of distinguishability is nice because of its relative simplicity in expression. However, this measure is not completely satisfactory in that it has no direct statistical inferential or information-theoretic meaning. Another measure of how distinct $p_0(b)$ and $p_1(b)$ are the actual Shannon information obtainable about the identity of the distribution.
$$I(p_0, p_1) = H(p) - \pi_0 H(p_0) - \pi_1 H(p_1)$$
where
$$H(q) = -\sum_b q(b) \log q(b)$$
is the Shannon entropy of a probability distribution q(b), $\pi_0$ and $\pi_1$ prior probabilities for s=0 and 1 respectively.

$p(b) = \pi_0 p_0(b) + \pi_1 p_0(b)$ is the overall prior probability for an outcome b. This measure is particularly appropriate for gauging Eve's measurement performance if her purpose is to identify or make an inference about a long string of quantum states prepared according to the distribution { $\pi_0$ and $\pi_1$ }. Finally, we consider the situation in which Eve bases her performance on the success of a guess about the identity of one single instance of Alice's prepared state. Suppose Eve obtains outcome b in her measurement. She can use this knowledge to update her probabilities or expectations about which state Alice really prepared. This is done formally via a use of Bayes' rule. Namely, after the measurement, her posterior expectation for the value of s is given by

$$p(s|b) = \frac{\pi_s p_s(b)}{p(b)}$$

Her probability of making an error will then be the minimum of p(0|b) and p(1|b). Averaging this over all possible outcomes gives the expected probability of error upon making the measurement $\hat{E}_b$:

$$P_e(p_0, p_1) = \sum_b \min\{\pi_0 p_0(b), \pi_1 p_1(b)\}$$

the smaller the overlap, the larger the Shannon mutual information, or the smaller the error probability of a guess, then, heuristically, the larger the "information" Eve has available about Alice's preparation. We may also retort to the concept of fidelity. It has several interesting properties despite its somewhat loose connection to statistical tests.

For instance, though it is necessarily bounded between 0 and 1, it equals unity if and only if the two quantum states in its arguments are identical.

$$F(\hat{\rho}_0^E, \hat{\rho}_1^E) = \langle \Psi_1^E | \hat{\rho}_0^E | \Psi_1^E \rangle$$

Two other important measures of distinguishability are: the Kolmogorov Distance and Bhattacharyya Coefficient [20].

The Kolmogorov distance between two density matrices is given by the relation

$$K(\rho_0, \rho_1) \stackrel{def}{=} \max_{\varepsilon \in M} K(p_o(\varepsilon), p_1(\varepsilon))$$

where the POVM $\varepsilon$ ranges over the set of all possible measurements $M$. The quantum mechanical relation is

$$K(\rho_0, \rho_1) = \frac{1}{2}\sum_{j=1}^{N} |\lambda_j| = \frac{1}{2} Tr |\rho_0 - \rho_1|$$

Two identical distributions have K=0, and two orthogonal distributions have K=1

The quantum Bhattacharyya coefficient can be expressed as

$$B(\rho_0, \rho_1) = Tr(\sqrt{\sqrt{\rho_0}\rho_1\sqrt{\rho_0}})$$

Two identical distributions have B=1, and two orthogonal distributions have B=0

Any attack on the classical channel will be in terms of information gain on the bit values in the final sifted key. In our scheme, the probability of the length (final key) being more than N/2 is more. The probability of Eve choosing the right sifting function is $NC_{N/2}$ based on the bases list.

The information gain can be obtained by calculating the difference in the Shannon entropy with (H1) and without (H2) ruling out the sifting functions that do not reproduce the final sifted key:

$$I = H1 - H2$$

where $H = -\sum_i p_i \log p_i$

$$I = \log_2(NC_{N/2}) - \log_2 \frac{(NC_{N/2})}{2^{N/2}} = \frac{N}{2}$$

But, the basis list is encrypted. So, Eve will have definite information with a probability.

Consequently, we have the net information gain

$$p = \frac{(NC_{N/2})}{2^{N/2}}$$

$$I = \frac{N}{2^{(N/2)+1}}$$

This quantized information gain is far lesser in comparison with the BB84 and its extensions. This value also suggests the number of bits to be refreshed so as to maintain the secrecy of the randomizer in case the bit values to be encoded remain the same for every run of the protocol. But this situation is very unlikely.

## 6. Extending Randomization to Entanglement based QKD

As a final step, we present an entanglement based protocol that includes the idea of encrypting data sent over public channel. The protocol extends the scheme proposed by Artur Ekert [3].

The key distribution is performed via a quantum channel which consists of a source that emits pairs of spin 1/2 particles in the singlet state

$$|\Psi\rangle = \frac{1}{\sqrt{2}}(|\uparrow\rangle|\downarrow\rangle - |\downarrow\rangle|\uparrow\rangle)$$

The Particles fly apart along the z-axis towards the two legitimate users of the channel, Alice and Bob, who, after the particles have separated, perform measurement and register spin components along one of three directions, given by unit vectors $\vec{a}_i$ and $\vec{b}_j$ (i; j = 1; 2; 3), respectively for Alice and Bob. For simplicity both $\vec{a}_i$ and $\vec{b}_j$ vectors lie in the x-y plane, perpendicular to the trajectory of the particles, and are characterized by azimuthal angles:
$\Phi_1^a = 0$, $\Phi_2^a = \frac{1}{4}\Pi$ $\Phi_3^a = \frac{1}{2}\Pi$,
and $\Phi_1^b = \frac{1}{4}\Pi$, $\Phi_2^b = \frac{1}{2}\Pi$, $\Phi_3^b = \frac{3}{4}\Pi$

Superscripts "a" and "b" refer to Alice's and Bob's analyzers respectively and the angle is measured from the vertical x-axis. The users choose the orientation of the analyzers randomly and independently for each pair of the incoming particles. The list of orientations chosen forms the bases list of Alice ($B_a$) and the observed bit values ($K_a$), the raw key. Similar is the case with Bob ($B_b$ and $K_b$). But raw key observed on either side of the channel may be of different versions.

If the choice of bases is the same for a particular pair, quantum mechanics predicts total anticorrelation of the results obtained by Alice and Bob:

$$E(\vec{a}_2, \vec{b}_1) = E(\vec{a}_3, \vec{b}_2) = -1$$

where the quantity

$$E(\vec{a}_i, \vec{b}_j) = P_{++}(\vec{a}_i, \vec{b}_j) + P_{--}(\vec{a}_i, \vec{b}_j) - P_{+-}(\vec{a}_i, \vec{b}_j) - P_{-+}(\vec{a}_i, \vec{b}_j)$$

is the correlation coefficient of the measurements performed by Alice along $\vec{a}_i$ and by Bob along $\vec{b}_j$. Here $P_{\pm\pm}(\vec{a}_i, \vec{b}_j)$ denotes the probability that result $\pm 1$ has been obtained along $\vec{a}_i$ and $\pm 1$ along $\vec{b}_j$.

One can define quantity $S$ composed of the correlation coefficients for which Alice and Bob used different bases for measurement

$$S = E(\vec{a}_1, \vec{b}_1) - E(\vec{a}_1, \vec{b}_3) + E(\vec{a}_3, \vec{b}_1) + E(\vec{a}_3, \vec{b}_3)$$

Quantum mechanics requires $S = -2\sqrt{2}$.

Both can now retort to an error correction procedure to be applied on the raw keys observed, to obtain an exact error corrected key (K). After the transmission has taken place, Alice and Bob can announce in public the orientations of the analyzers they have chosen for each particular measurement XORed with K. Note that the public announcement of the bases lists has been encrypted. The encrypted bases list is decrypted by performing XOR with K on each side. They divide the measurements into two separate groups: a first group for which they used different bases, and a second group for which they used the same bases. They discard all measurements in which either or both of them failed to register a particle at all. Further, any eavesdropping during quantum transmission can be detected by computing the correlation coefficient S on unequal bases.

## 7. Summary and Conclusion

At the very outset, the protocol suggested by us is essentially a reordering of the operations in the standard BB84, with the additional security of encrypted public transmission about which Eve will attain practically lesser information. The EPP protocols whose security considerations are well analyzed can be reduced to the present form of our protocol. The extension to entanglement based scheme results in a more secure protocol whose practical applicability needs to be analyzed. An improvement of the protocol can be developed with consideration to details concerning apparatus used for transmitting quantum information, for better generalization. The future enhancements involve statistical analysis of the performance of the protocol using QuCrypt.